\documentclass[12pt,preprint]{aastex}

\begin{document}

\title {A FLARED, ORBITING, DUSTY DISK AROUND HD 233517}

\author{M. Jura} 
\affil{Department of Physics and Astronomy, University of California,
    Los Angeles CA 90095-1562; jura@clotho.astro.ucla.edu}

\begin{abstract}
We find that the infrared excess around HD 233517, a first ascent red giant, can be naturally explained if the star possesses an orbiting, flared dusty disk.  We estimate that the outer radius of this disk is ${\sim}$45 AU and that the total mass within the disk is ${\sim}$0.01 M$_{\odot}$.   We speculate that
this disk is the result of the engulfment of a low mass companion star that occurred when HD 233517 became a red giant.  
  
\end{abstract}

\section{INTRODUCTION}
The  evolution of dust disks into large solid bodies such as asteroids, planets and comets is of central importance in astronomy.  While it is usually assumed that such disks are primordial and associated with young stars,  some binary systems such as the Red Rectangle
apparently create orbiting  reservoirs of gas and dust during their post main sequence evolution (see Waters et al. 1993, Jura \& Kahane 1999).  To date, there are only a few  good candidate systems  for orbiting
dusty circumstellar material around evolved stars, and therefore every example of this phenomenon is worth studying.  Here, we propose that the puzzling infrared
excess around HD 233517, a first ascent red giant, can be naturally understood as resulting from a flared, orbiting disk.

 Attention was originally drawn to HD 233517 because it is a K type star (m$_{V}$ = 9.72 mag) that is an IRAS source with an unusually large fractional infrared
excess ($L_{IR}/L_{*}$ ${\sim}$ 0.06, Sylvester, Dunkin \& Barlow 2001) produced by  cold grains ($T$ ${\sim}$ 100 K), and it was thought to be
an example of the Vega phenomenon -- a main sequence star with dust (Walker \& Wolstencroft 1988).   
However, detailed 
spectroscopic studies strongly suggest that HD 233517 is a first ascent red giant with a luminosity near 100 L$_{\odot}$
(Fekel et al. 1996, Balachandran et al. 2000, Zuckerman 2001).  While infrared excesses are common around    
 second ascent red giants on the Asymptotic Giant Branch  with luminosities in excess of 1000 L$_{\odot}$
(see, for example, Habing 1996),   relatively few  first ascent red giant stars
with their lower luminosities  have detectable infrared excesses  (Judge, Jordan \& Rowan-Robinson 1987, Zuckerman, Kim \& Liu 1995). 
 HD 233517 has a uniquely high value of $L_{IR}/L_{*}$ among first ascent red giants where this ratio is generally less than 10$^{-3}$ (see Zuckerman et al. 1995).  It is unlikely that HD 233517 is a pre-main sequence star since
it does not lie near any known region of star formation and since it lies
far from the location of young stars in the H-R diagram  (Fekel et al. 1996, Balachandran et al. 2000).  Also, HD 233517 appears to be similar to
a few other K giants  which rotate rapidly, have strong lithium lines, a
detectable far-infrared excess and are definitely post main sequence since
at least in the case of PDS 365, the carbon isotope ratio, $^{12}$C/$^{13}$C, is approximately 12 (Drake et al. 2002).  

  Three models to explain infrared excesses around first ascent red giants are (1) dust is being produced by mass loss, (2) nearby interstellar dust is illuminated accidentally by the red giant, and (3) the  dust is orbiting (see Jura 1999, Kalas et al. 2002).  The fraction of red giants that
display true dust excesses is uncertain and controversial (Jura 1990, Plets et al. 1997, Jura 1999, Kim, Zuckerman \& Silverstone 2001).  
In any case, since ground-based  imaging shows that the 10 ${\mu}$m excess is   physically associated with HD 233517 (Skinner et al. 1995, Fisher et al. 2000), at least for this particular star, the dust producing the infrared excess  either is orbiting or is expanding away from the star.  For some other stars, the infrared excess has been resolved with ISO and in these objects, the dust is probably not orbiting (Kim et al. 2001).  

There are difficulties with the model that HD 233517 is currently losing enough
mass to produce the observed infrared excess. As with other first ascent
red giants with dust, the infrared spectrum of HD 233517 peaks near
60 ${\mu}$m.  This spectral energy distribution is characteristic of cool material and is very different from that associated with
a continuous mass loss rate where there is a substantial amount of dust
near a temperature of 1000 K (see, for example, Sopka et al. 1985).  One possibility
is that the mass loss rate from HD 233517 is episodic and  currently the star is not expelling much matter.  However, as noted by Jura (1999), the characteristic time scale for the dust to expand  around a first ascent red
giant to its inferred location may be less than 20 years.  There is no evidence for recent infrared variability of this star, although the star does exhibit a full amplitude optical variation of 0.02 mag with a 47.9 day period  which may be caused by star spots rotating in and out of the view (Balachandran et al. 2000).   

Since the dust around HD 233517 may not be carried in a wind, we consider models where the dust is orbiting the star.
One scenario is that the dust around HD 233517 is simply ``left over"
from the main sequence phase.  However, a major difficulty with this model
is that the fractional infrared excess around HD 233517, $L_{IR}/L_{*}$, of 0.06  is much larger than that found
for even the ``dustiest'' main sequence stars such as ${\beta}$ Pic and HR 4796 where $L_{IR}/L_{*}$ is 2-4 ${\times}$ 10$^{-3}$ (see,
for example, Zuckerman 2001).  

Here, we  speculate that HD 233517 had a low mass companion which was engulfed 
when it became a red giant.  Although uncertain, the engulfment of this companion could have led to the ejection of an equatorial ring of orbiting material (see Taam \& Sandquist 2000, Spruit \& Taam 2001).   As discussed by Pringle (1991), after this ring is created, the matter expands under the action of tidal torques which transfer angular momentum from
the binary into the circumstellar material which thus evolves into an extended disk.    Although not required in Pringle's model, it is possible that in this dense
equatorial ring, dust grains formed.
We suggest that the structure of the circumstellar system is described by the models of  Chiang \& Goldreich (1997) for passive, orbiting disks which flare in response
to the illuminating radiation.  
Below, we present details
of this  idealized model.      

\section{THE DISK AND THE STAR}

\subsection{Stellar parameters}
First, we
list our assumed parameters for the star.  
 Following
Balachandran et al. (2000), we adopt a distance, $D_{*}$, of 620 pc, an effective temperature, $T_{*}$, of 4475 K a radius, $R_{*}$, of 1.1 ${\times}$ 10$^{12}$ cm and a luminosity of 100 L$_{\odot}$.  We assume that the mass of the star, $M_{*}$,
is approximately 1 M$_{\odot}$, consistent with the star's location on the
H-R diagram and measured surface gravity (Balachandran et al. 2000).  The rotational velocity of the star is $v\,sin\,i$ = 17.6 km s$^{-1}$ which implies if the star rotates as a solid body and if the moment of inertia of the star is $0.1\,M_{*}\,R_{*}^{2}$  (taken from Schwarzschild 1958) that the angular momentum of the star is
at least 4 ${\times}$ 10$^{50}$ g cm$^{2}$ s$^{-1}$.  
\subsection{Predicted Infrared Emission}

     Dust is required to explain the emission from HD 233517 for ${\lambda}$ ${\geq}$ 10 ${\mu}$m (see Skinner et al. 1995).  Here, we use the model of a opaque, flared disk which is locally in vertical hydrostatic equilibrium (Chiang \& Goldreich 1997).   In this model, the
temperature, $T_{disk}$, as a function of distance from the star, $R$, is
 (see Jura et al. 2002):
\begin{equation}
T_{disk}\;=\;\left(\frac{1}{7}\right)^{2/7}\left(\frac{R_{*}}{R}\right)^{3/7}\,\left(\frac{2k_{B}T_{*}R_{*}}{GM_{*}{\mu}}\right)^{1/7}\;T_{*}
\end{equation}     
where $k_{B}$ is Boltzmann's constant, $G$ is the gravitational constant and
${\mu}$ is the mean molecular weight of the gas.  Here, we assume that
the gas is primarily H$_{2}$ with [He]/[H$_{2}$] = 0.2 so that ${\mu}$ = 3.9 ${\times}$ 10$^{-24}$ g.  In view of the many uncertainties, and to be consistent with inferences for disks around pre-main sequence stars (Chiang et al. 2001), we assume a simple model where the half-thickness of the vertically-isothermal disk is taken equal to the parameter, $h$, in Jura et al. (2002) which is a factor of ${\sqrt{2}}$ larger than $h$ in Chiang \& Goldreich (1997), that characterizes the Gaussian density distribution above the plane.   

Since the disk is opaque, we can
determine the flux from the source, $F_{\nu}$, by:
\begin{equation}
F_{\nu}\;=\;\frac{2\,{\pi}\,cos\,i}{D_{*}^{2}}\;{\int}_{0}^{R_{out}}\;B_{\nu}(T_{disk})\,R\,dR
\end{equation}
where $R_{out}$ denotes the outer boundary of the disk where the temperature is $T_{out}$.  For simplicity, we assume a sharp outer boundary although  an exponential decay would be more realistic (see Jura et al. 2002). It is convenient
to introduce the dimensionless parameter, $x$:
\begin{equation}
x\;=\;\frac{h{\nu}}{k_{B}T}
\end{equation}
From equations (1) and (2):
\begin{equation}
F_{\nu}\;=\;\frac{28{\pi}}{3}\,\frac{cos\,i\,R_{*}^{2}}{D_{*}^{2}}\,\left(\frac{k_{B}T_{*}}{h{\nu}}\right)^{5/3}\frac{(k_{B}T_{*})^{3}}{(hc)^{2}}\left(\frac{2\,k_{B}T_{*}R_{*}}{49\,G\,M_{*}\,{\mu}}\right)^{2/3}\,{\int}_{0}^{x_{out}}\frac{x^{11/3}}{e^{x}\;-\;1}\,dx
\end{equation}
At high frequencies with $x_{out}$ $>$ 10, the observed flux is insensitive to the outer boundary condition since the integral in equation (4)  is approximately 15. The
spectrum is predicted to vary as ${\nu}^{-5/3}$.   At low frequencies where $x_{out}$ $<$2, we may re-write equation (4) to find that:
\begin{equation}
F_{\nu}\;=\;\frac{28{\pi}}{11}\,\frac{cos\,i\,R_{*}^{2}}{{\lambda}^{2}\,D_{*}^{2}}\,\left(\frac{T_{*}}{T_{out}}\right)^{11/3}\left(\frac{2\,k_{B}T_{*}R_{*}}{49\,G\,M_{*}\,{\mu}}\right)^{2/3}\,k_{B}\,T_{*}
\end{equation}
If the disk is opaque, then at low frequencies, F$_{\nu}$ varies as ${\nu}^{2}$.

For the data in the IRAS Faint Source Catalog, between 12 ${\mu}$m and 60 ${\mu}$m, F$_{\nu}$ varies as ${\nu}^{-1.70}$, in good agreement with the prediction that
F$_{\nu}$ should vary as ${\nu}^{-1.67}$.  Furthermore, 
with $x_{out}$ $>>$ 1 and cos $i$ = 0$^{\circ}$,  from equation (4), the predicted value of F$_{\nu}$(25 ${\mu}$m)  is 3.9 Jy which agrees
with the observed value of  3.4 Jy.  Thus, without any 
``tweaking", the predictions  are close enough to the data that the model
should be taken seriously.  

Near 60 ${\mu}$m, the slope of the spectral energy distribution turns over. 
In the flared disk model, the peak in the spectral energy distribution is governed by the outer boundary of the disk.  As shown in Figure 1, we have found that a model with  $T_{out}$ = 70 K, which occurs at $R_{out}$ = 45 AU, can reproduce much of the infrared data.    
However, at 1.35 mm, much more flux is predicted than the observed value of 2.6 ${\pm}$ 1.0, which is conservatively interpreted as a  3${\sigma}$ upper limit of 3 mJy 
by Sylvester et al. (2001).  A possible resolution of this
discrepancy  is discussed below.
\subsection{Predicted Millimeter Continuum Emission}
   The flared disk model drastically overestimates the flux at 1.35 mm.  
It is observed that between 1.35 mm and 100 ${\mu}$m, F$_{\nu}$ varies as ${\nu}^{2.8}$ which suggests that at least at the longer wavelength, the system is optically thin.
If the circumstellar dust particles are smaller than 100 ${\mu}$m in diameter, then the opacity at 1.35 mm is probably considerably smaller than the opacity at 100 ${\mu}$m.  
Although the composition and size distribution of the dust particles are unknown; we adopt  parameters
representative of circumstellar ``silicates" from  Ossenkopf, Henning \& Mathis (1992) and Pollack et al. (1994).  
Thus, we take ${\chi}_{\nu}$(100 ${\mu}$m) = 25 cm$^{2}$ g$^{-1}$, and for ${\lambda}$ $>$ 100 ${\mu}$m, we assume that ${\chi}_{\nu}$ scales as ${\nu}^{2}$ so that ${\chi}_{\nu}$(1.35 mm) = 0.14 cm$^{2}$ g$^{-1}$.    

To calculate the flux for the optically thin regime, we need to describe
the mass distribution.      
If ${\Sigma}$ denotes the total dust surface density of the disk, then
we approximate the model of Pringle (1991) for a circumbinary disk by assuming
that all the matter is confined within angular radius ${\phi}_{out}$, where ${\phi}$ = $R/D_{*}$,  and
that within this region, 
\begin{equation}
{\Sigma}\;=\;{\Sigma}_{0}\,\left(\frac{{\phi}_{out}}{{\phi}}\right)^{3/2}
\end{equation}  
With this description, if $M_{dust}$ denotes the mass of dust in the disk, then:
\begin{equation}
{\Sigma}_{out}\;=\;\frac{M_{dust}}{4\,{\pi}\,R_{out}^{2}}
\end{equation}

The flux from the disk, $F_{disk}$, is:
\begin{equation}
F_{disk}\;=\;cos\,i\,{\int}_{0}^{{\phi}_{out}}\,2\,{\pi}\,{\phi}\,d{\phi}\,
B_{\nu}(T)\,\left(1\,-\,exp^{-{\chi}_{\nu}\,{\Sigma}_{dust}({\phi})/cos\,i}\right)
\end{equation}
We evaluate equation (8)  for the condition that at $R$ = $R_{out}$ (or 45 AU), ${\Sigma}_{out}$ = 0.04 g cm$^{-2}$.  As discussed below,   this value of ${\Sigma}_{out}$ is chosen  so that the disk is
opaque at 100 ${\mu}$m at its outermost region. The results are
shown in Figure 1.  With the revised model,  we see 
 that the predicted flux at 1.35 mm
is consistent with the upper limit reported by Sylvester et al (2001).

\subsection{Predicted CO Emission}
The flared disk model requires that gas  be in hydrostatic equilibrium vertical to the orbiting plane of the system.  We can 
estimate the expected amount of CO emission from HD 233517
if the gas and grain temperatures are equal to each other.  First,
\begin{equation}
T\;=\;T_{out}\left(\frac{{\phi}_{out}}{{\phi}}\right)^{3/7}
\end{equation}
 If the gas is optically thick, then
 in a telescope of beam size, ${\Omega}_{tel}$, 
the observed brightness temperature, $T_{obs}$ is
\begin{equation}
T_{obs}\;=\;\,cos\,i\,{\int}_{0}^{{\phi}_{out}}\;\frac{2{\pi}\,{\phi}\,T({\phi})\,d{\phi}}{{\Omega}_{tel}}\;=\;\frac{14\,{\pi}}{11}\frac{T_{out}{\phi}_{out}^{2}\,cos\,i}{{\Omega}_{tel}}
\end{equation}
Within the 10{\farcs}5 beam of the IRAM telescope,
Jura \& Kahane (1999) placed a 1${\sigma}$  upper limit to the CO J 2${\rightarrow}$1 emission
line of 29 mK.  In our model, ${\phi}_{out}$ =  0{\farcs}073, and the predicted brightness temperature in this line is 17 mK.  Thus the model is consistent  with  the observed upper limit to the
CO line emission.
\subsection{Disk Mass and Angular Momentum}

 We derive a minimum mass of the disk from
the criterion that our model requires the disk to be opaque even at $R$ = $R_{out}$ at 100 ${\mu}$m or ${\Sigma}_{out}\,{\chi}_{\nu}$(100 ${\mu}$m) $>$ 1.  With
${\chi}_{\nu}$(100 ${\mu}$m) = 25 cm$^{2}$ g$^{-1}$, this implies that
${\Sigma}_{out}$ ${\geq}$ 0.04 cm$^{2}$ g$^{-1}$.  Therefore, from equation
(7), the mass of the dust in the disk is 2 ${\times}$ 10$^{29}$ g.
If the  gas to dust ratio is 100,  the total disk mass is ${\sim}$0.01 M$_{\odot}$ 

Another important parameter is the disk angular momentum, $J$.
With a sharp outer boundary and the surface density variation given by equation (6), we may write:
\begin{equation}
J\;=\;\frac{M_{disk}}{2}\,\sqrt{G\,M_{*}\,R_{out}}
\end{equation}
We therefore estimate a total angular momentum of the disk of 3 ${\times}$ 10$^{51}$ g cm$^{2}$ s$^{-1}$, which is larger than  the total angular momentum of the star estimated
above to be ${\geq}$ 4 ${\times}$ 10$^{50}$ g cm$^{2}$ s$^{-1}$.
\section{DISCUSSION}
We have found that the infrared fluxes from HD 233517 are naturally explained with the simple flared disk model.  The origin of this disk is uncertain.
One possible model to account both for the creation of a disk and the
unusually high rotation speed of the red giant is that while on the main sequence,  HD 233517 
had a companion at an orbital separation  of
about 1 ${\times}$ 10$^{12}$ cm (see Soker, Livio \& Harpaz 1984).  We speculate that when HD 233517 became a red giant,
it engulfed this companion with the consequences of both creating an orbiting
ring of circumbinary matter and spinning up the primary star.  With the parameters derived above, the total angular momentum of HD 233517 and its associated disk is
3 ${\times}$ 10$^{51}$ g cm$^{2}$ s$^{-1}$.  This amount of angular momentum
could have been supplied by a companion star of 0.12 M$_{\odot}$ in  a circular orbit of radius 1.1 ${\times}$ 10$^{12}$ cm.

HD 233517 is a lithium-rich star (Fekel et al. 1996, Balachandran et al. 2000).  While single-star models might explain this result (de la Reza et al. 1997, Jasniewicz et al. 1999, Charbonnel \& Balachandran 2000), in our picture, this lithium enrichment results because the system was a binary. The absence of $^{6}$Li argues against the simple idea that incorporation of lithium from a planet or low mass companion into the envelope of the star (Balachandran et al. 2000) can explain the observed abundance of this element, but the capture of a companion might have led to an episode that 
 stimulated interior mixing of lithium to the surface (Denissenkov \& Weiss 2000). 

With a value of $L_{IR}/L_{*}$ $>$ 10$^{-2}$, HD 233517 is a distinctive red giant.  In the sample of 
Zuckerman et al. (1995) of more than 40,000 of the nearest luminosity class III giants, the highest value of $L_{IR}/L_{*}$ is 10$^{-3}$.  Although the statistics are poor, perhaps only
10$^{-4}$ of all first ascent red giants have as much circumstellar opacity as does HD 233517.  
We suggest that
HD 233517 is unusual for two reasons.  First, according to Duqennoy \& Mayor
(1991), only ${\sim}$2\% of all solar-mass main sequence stars have detectable   companions with orbital periods near 10 days.  Such binaries are the kinds of systems that can evolve into  a red giant with a substantial disk as is found
around HD 233517.  Furthermore, once the disk is formed, it must have a limited
lifetime.  
The massive dusty disk
around HD 233517 may evolve in a fashion somewhat like disks around pre-main sequence stars.   For 
A type stars which have luminosities somwhat less than that of HD 233517,  the 
lifetime of the disk as a system with $L_{IR}/L_{*}$ $>$ 10$^{-2}$ is near 3 ${\times}$ 10$^{6}$ years,
although some small amount of dust could linger for as long as 4 ${\times}$ 10$^{8}$ years (see Habing et al. 1999, Spangler et al. 2001).  Since the lifetime of a 1 M$_{\odot}$ red giant as it evolves from 10 L$_{\odot}$ to its maximum luminosity of 2700 L$_{\odot}$ is ${\sim}$2 ${\times}$ 10$^{8}$ years (Girardi et al. 2000),  only ${\sim}$  1\% of the disks around red giants might
have $L_{IR}/L_{*}$ as large as 10$^{-2}$.  Given these two
considerations, it is not too surprising that  a system like
HD 233517 occurs among first ascent red giants with a frequency of perhaps 10$^{-4}$. A plausible explanation for the inferred
decrease of $L_{IR}/L_{*}$ with time is that the dust particles coalesce
into larger bodies.
 
\section{CONCLUSIONS}
We have modeled the currently available infrared data for HD 233517 and find
that the observations can be naturally explained 
by a flared, orbiting disk of mass  ${\sim}$0.01 M$_{\odot}$ and   outer radius
of  ${\sim}$45 AU.  We speculate that this disk was
created by the engulfment of a low mass stellar companion. 

This work has been partly supported by NASA.

\newpage
\begin{figure}
\epsscale{1}
\plotone{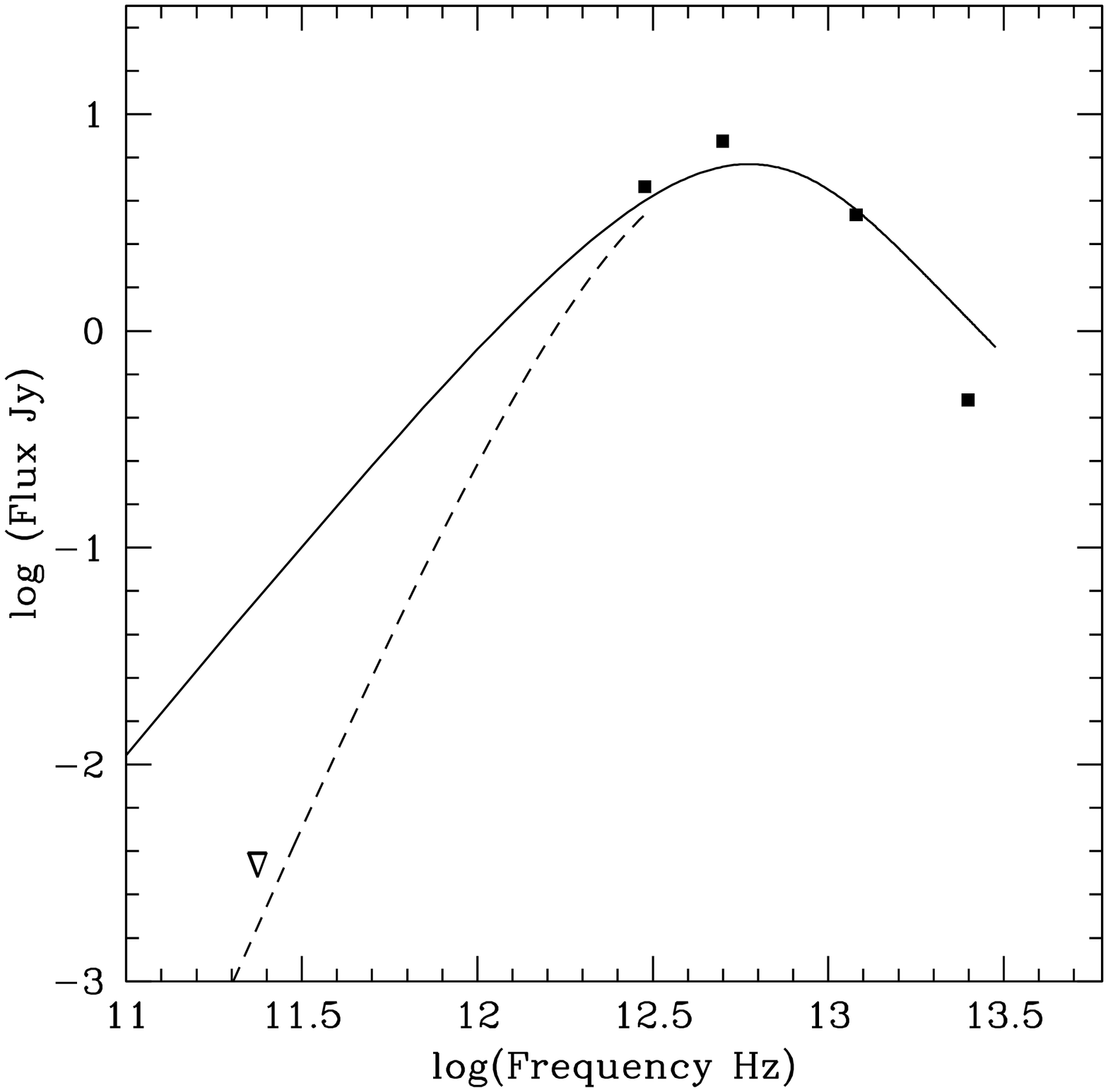}
\caption {A plot of the model spectral energy distribution for HD 233517 compared to observations.   The squares show the data from IRAS; the upper limit to the flux at 1.35 mm  (Sylvester et al. 2001) is shown
as the upside triangle.  The solid line shows the
model for the opaque, flared disk;  the dashed line shows the modification to this model if the outer portion of the disk becomes
optically thin at ${\lambda}$ $>$ 100 ${\mu}$m as described in the text.}
\end{figure}

\begin{thebibliography}{}
\bibitem{balachandran00} Balachandran, S. C., Fekel, F. C., Henry, G. W. \& Uitenbroek, H. 2000, \apj, 542, 978
\bibitem{charbonnel00} Charbonnel, C. \& Balachandran, S. C. 2000, \aap, 359, 563
\bibitem{chiang97} Chiang, E. I. \& Goldreich, P. 1997, \apj, 490, 368
\bibitem{chiang01} Chiang, E. I., Joung, M. K., Creech-Eakman, M. J., Qoi, C., Kessler, J. E., Blake, G. A. \& van Dishoeck, E. F. 2001, \apj, 547, 1077
\bibitem{delareza97} de la Reza, R. Drake, N. A., Da Silva, L., Torres, C. A. O. \& Martin, E. L. 1997, \apj, 482, L77
\bibitem{denissenkov00} Denissenkov, P. A. \& Weiss, A. 2000, \aap, 358, L49
\bibitem{drake02} Drake, N., de la Reza, R., da Silva, L. \& Lambert, D. L. 2002, \aj, 123, 2703
\bibitem{duquennoy91} Duquennoy, A. \& Mayor, M. 1991, \aap, 248, 485
\bibitem{fekel96} Fekel, F. C., Webb, R. A., White, R. J. \& Zuckerman, B. 1996, \apj, 462, L95
\bibitem{fisher00} Fisher, R. S., Telesco, C. M., Pina, R. K. \& Knacke, R. F. 2000, in Disks, Planetesimals, and Planets, ASP Conf Ser. v. 219, ed. F. Garzon, C. Eiroa, D. de Sinter, T. J. Mahoney, San Francisco ASP, 369
\bibitem{girardi00} Girardi, L., Bressan, A., Bertelli, G. \& Chiosi, C. 2000, \aaps, 141, 371
\bibitem{habing96} Habing, H. J. 1996, A\&Ap Rev., 7, 97
\bibitem{habing99} Habing, H. J., Dominik, C., Jourdan de Muizon, M., Kessler, M. F., Laureijs, R. J., Leech, K., Metcalfe, L., Salama, A., Siebermorgen, R. \& Trams, N. 1999, Nature, 401, 456
\bibitem{jasniewicz99} Jasniewicz, G., Parthasarathy, M., de Laverny, P. \& Thevenin, F. 1999, \aap, 342, 831
\bibitem{jura90} Jura, M. 1990, \apj, 365, 317
\bibitem{jura99} Jura, M. 1999, \apj, 515, 706
\bibitem{jura02} Jura, M., Chen, C. \& Plavchan, P. 2002, \apj, 574, 963
\bibitem{jura99} Jura, M. \& Kahane, C. 1999, \apj, 521, 302
\bibitem{judge87} Judge, P. G., Jordan, C. \& Rowan-Robinson, M. 1987, \mnras, 224, 93
\bibitem{kalas02} Kalas, P., Graham, J. R., Beckwith, S. V. W., Jewitt, D. C. \& Lloyd, J. P. 2002, \apj, 567, 999
\bibitem{kim01} Kim, S. S., Zuckerman, B. \& Silverstone, M. 2001, \apj, 550, 1000
\bibitem{ossenkopf92}Ossenkopf, V., Henning, Th. \& Mathis, J. S. 1992, \aap, 261, 567
\bibitem{plets97} Plets, H., Waelkens, C., Oudmaijer, R. D. \& Waters, L. B. F. M. 1997, \aap, 323, 513.  
\bibitem{pollack94} Pollack, J. B., Hollenbach, D., Beckwith, S., Simonelli, D. P., Roush, T. \& Fong, W. 1994, \apj, 421, 615
\bibitem{pringle91} Pringle, J. E. 1991, \mnras, 248, 754
\bibitem{schwarzschild58} Schwarzschild, M. 1958, in Structure and Evolution of the Stars, Dover Publications: New York
\bibitem{skinner95} Skinner, C. J., Sylvester, R. J., Graham, J. R., Barlow, M. J., Meixner, M., Keto, E., Arens, J. F. \& Jernigan, G. 1995, \apj, 444, 861
\bibitem{soker84} Soker, N., Livio, M. \& Harpaz, A. 1984, \mnras, 210, 189
\bibitem{sopka85} Sopka, R. J., Hildebrand, R., Jaffe, D. T., Gatley, I., Roellig, T., Werner, M., Jura, M. \& Zuckerman, B. 1985, \apj, 294, 242
\bibitem{spangler01} Spangler, C., Sargent, A. I., Silverstone, M. D., Becklin, E. E. \& Zuckerman, B. 2001, \apj, 555, 932
\bibitem{spruitt01} Spruit, H. C. \& Taam, R. E. 2001, \apj, 548, 900
\bibitem{sylvester01} Sylvester, R. J., Dunkin, S. K. \& Barlow, M. J. 2001, 
\mnras, 327, 133
\bibitem{taam20} Taam, R. E. \& Sandquist, E. L.  2000, ARA\&Ap, 38, 113
\bibitem{walker88} Walker, H. J. \& Wolstencroft, R. D. 1988, \pasp, 100, 1509
\bibitem{waters93} Waters, L. B. F. M., Waelkens, C., Mayor, M. \& Trams, N. R. 1993, \aap, 269, 242
\bibitem{zuckerman95} Zuckerman, B., Kim, S. S. \& Liu, T. 1995, \apj, 446, L79
\bibitem{zuckerman01} Zuckerman, B. 2001, ARA\&Ap, 39, 549
\end{thebibliography}
\end{document}